# Analysis of Various Symbol Detection Techniques in Multiple-Input Multiple-Output System (MIMO)

Shuchi Jani[1], Shrikrishan Yadav[2], B. L. Pal[3]

[1]CSE Dept, SITE, Nathdwara, India

janishuchi@gmail.com

[2]CSE Dept, PAHER University, Udaipur, India

Shrikrishanyadav77@gmail.com

[3]CSE Dept, MEWAR University, Chittorgarh, India

Contact2bl@rediffmail.com

## ABSTRACT

*Wireless communication is the fastest growing area of the communication industry. To keep swiftness with the indefinite increase in customers' demands and expectations, and the market competition among companies for the services offered ,there is need for higher data rate along with reliable communication at low cost so that the applications can reach all. Until now, many technical challenges remains in designing robust and fast wireless systems that deliver the performance necessary to support emerging applications, due to the fact that wireless channel is frequency selective, power-limited, susceptible to noise and interference. Demand for high data rate and increasing applications offered by a wireless device calls for an effective method. Due to limit on the available bandwidth, there is a need for exploiting the available bandwidth in a way so that we get maximum advantage. Multiple-Input Multiple-Output system does exactly this thing by multiplying the data rate without any expansion in the bandwidth. This system utilizes the spatial diversity property of the multi channel system. The reliable transmission requires symbols to be effectively recovered at the receiving end. V-BLAST detection technique is employed for this purpose. This paper depicted the advantages of using multiple antennas by exploiting signal diversity offered by multipath effect and the system offers high spectral efficiency.*

## KEYWORDS

*Wireless Communication System, Antenna, Multiplexing, MIMO, ML, QAM, MMSE, V-BLAST*

## 1. INTRODUCTION

Multiple-input multiple-output is a multiple antenna technology for communication in wireless systems. Multiple antennas are used at both the source (transmitter) and the destination (receiver). The antennas at each end of the communications system are combined to minimize errors and optimize data speed. The radio wave propagating through the wireless channel undergo transmit power dissipation (path loss) and shadowing caused by obstacles on the course from transmitter to receiver that attenuate signal power through absorption, reflection, scattering and diffraction. Constructive and destructive addition of different multipath components is even introduced by the wireless channel to cause the fading effect, which is generally considered as a serious impairment to the wireless channel. The MIMO symbol detection methods are observed under frequency flat fading AWGN channel condition and by employing the Quadrature amplitude modulation (QAM) method.

The various symbol detection techniques are compared to observe their behavior under AWGN channel condition. Maximum likelihood (ML) symbol detection method gives the best performance but because of its high complexity it can't be used. Sphere decoder reduces the complexity to some extent providing similar performance as ML estimate. The other methods used are Zero forcing and Minimum mean square estimation (MMSE). These two methods are used successively for interference cancellations improve performance to large extent along with reduction in the cost.

The data rate of a communication system can be increased by using the following techniques [5]

• Increasing the transmitter's effective isotropic radiated power (EIRP) or reducing system losses (SNR is increased)

• Increasing the available channel bandwidth

• Utilizing the communication resource more efficiently

Due to the high cost involved in first two we go for improving the third factor.

To obtain these spectral efficiency improvements, we would often need knowledge of the channel condition, which is represented by the channel matrix. The cost of the performance enhancements achieved through MIMO techniques comes from deploying multiple antennas, the space and power requirements to install these extra antennas and the additional computing complexity to process multidimensional signals.

In order to investigate the channel model we describe MIMO channel at certain time n.We consider the V-BLAST system with $N_t$ transmit and $N_r$ receive antenna. The transmitted symbol vector is given as $x[n] = [x_1.....x_{Nt}]^T$ ,and the received vector is

$$y[n] = Hx[n] + w[n] \tag{1}$$

In, $W[n] = [W1....W_{Nr}]$ T represents noise. The channel matrix H is

$$H = \begin{bmatrix} h_{1,1} & h_{1,2} & \cdots & h_{1,Nt} \\ h_{2,1} & h_{2,2} & \cdots & h_{2,Nt} \\ \vdots & \vdots & \ddots & \vdots \\ h_{Nr,1} & h_{Nr,1} & \cdots & h_{Nr,Nt} \end{bmatrix} \tag{2}$$

Here $h_{i,j}$ represents the complex gain from $j^{th}$ transmitting antenna to the $i^{th}$ receiving antenna.

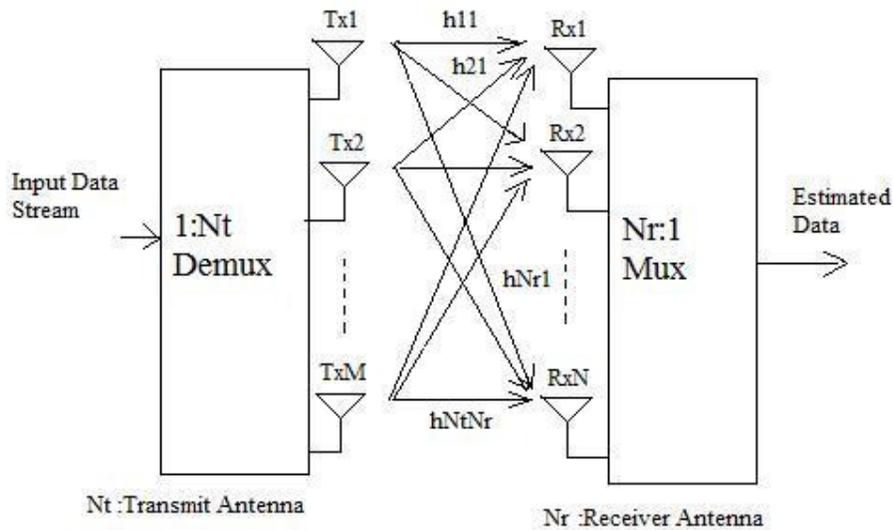

Figure 1. Block diagram of MIMO system

## 2. RELATED WORK

The evolution of MIMO system started from the work of Winters [1], Foschini and Gans [2], and Telatar [3] [4]. In the recent years, multiple-input multiple-output (MIMO) communication over multiple antenna channels has attracted the attention of many researchers. In [7], a transmission technique called V-BLAST is introduced for high data rate communication over point to point MIMO channels. In MIMO system symbols are transmitted independently over different transmit antennas. Most of the earlier decoding methods for BLAST (such as nulling and cancelling, zero forcing and MMSE-type methods) cannot achieve the maximum receive diversity which is equal to the number of receive antennas. The Maximum likelihood (ML) estimation outperforms all the methods but has high computational complexity. Efforts are made to reach the performance of ML estimation while reducing complexity. Now days, in MIMO system transmitted data is coded to increase the spatial diversity. Space time frame coding is one such method in which convolution coding is used for symbol coding and decoding is done using Viterbi decoder. STTC also adds coding gain along with improving spectral diversity which improves the performance of the system.

## 3. MODULATION METHODS

Modulation is the process where information is added to the radio carrier. There are different types of modulation methods but out which, digital modulation schemes are the obvious choices for the future wireless communication system when data services such as wireless multimedia are to be supported. Digital modulation improves spectral efficiency because digital signal are more robust against channel impairments. To achieve high spectral efficiency the modulation scheme with high bandwidth efficiency in units of bits per seconds per hertz of bandwidth must be selected.

The choice of a modulation scheme depends upon the following properties:

- Compact power density spectrum: To minimize the effect of adjacent channel interference it is desirable that the power radiated into the adjacent channel be 60-80 dB below that in the desired channel. Hence, modulation scheme with a narrow main lobe and fast roll off of side lobes are desirable.

- Good bit error rate performance: A System with low bit error rate probability even in the presence of co channel interference, adjacent channel interference, thermal noise and other channel impairments such as fading and inter symbol interference (ISI).

- Envelope properties: The input signal should have a relatively constant envelope to prevent the re-growth of spectral side lobes during nonlinear amplification.

## 4. SYMBOL DETECTION METHODS

V-BLAST is architecture for realizing very high data rate over a rich scattering wireless channel. It is a multi-layer symbol detection scheme which detects symbols transmitted at different transmit antennas successively in a certain data independent order. BLAST is an extraordinarily bandwidth efficient approach. In flat fading MIMO channels having multiple transmit and receive antennas were shown to offer relatively huge spectral efficiencies compared to SISO (Single Input and Single Output) channels [6][7]. Capacity increases linearly with the number of transmit antennas as long as the number of receive antennas is greater than or equal to the number of transmit antennas. To achieve this capacity, Diagonal BLAST was proposed by Foschini [7]. This scheme utilizes multi-element antenna arrays at both ends of wireless link. However, the complexities of D-BLAST implementation lead to V-BLAST which is a modified version of BLAST [8].

The V-BLAST detector decodes the sub-streams using a sequence of nulling and cancellation steps. An estimate of strongest transmitted signal is obtained by nulling out all the weaker transmit signal using the zero forcing or MMSE criterion. Then this strongest signal component is subtracted from the received signal, and we proceed to decode the next strongest signal of the remaining transmitted signal and this loop goes on until all symbols are detected. The BLAST detection scheme was based on a successive interference cancellation [8] [9] [10]. A parallel interference cancellation scheme was also proposed later [11]. BLAST detectors including both SIC and PIC suffer from the error propagation problem, so that they lead to the poor energy efficiency which can be improved if the previously detected layers were perfectly cancelled because the following layers depend highly on the result of the previous detected signals. The error propagation problem of BLAST detectors can be reduced with channel coding and interleaving [12] [13].

The detection strategy is one of the prime criteria to determine the effectiveness of a communication system. There are different methods such as Zero forcing (ZF), Minimum Mean Square Estimation (MMSE), V-BLAST/ZF, V-BLAST/MMSE and Maximum Likelihood (ML) which are analyzed and compare the performance of these systems. The best performance is obtained by using ML estimate, but it has a fault .The computational complexity is very high and it increases with the increase in number of transmitting and receiving antennas and the constellation size. Other alternative were then thought of which reduces complexity and gives performance close to ML estimation scheme.

Some assumptions are made in different symbol detections algorithms such as:

1. Frequency flat AWGN channel condition

2. Input signal and noise are uncorrelated

3. Number of receiving antenna is greater than number of transmitting antennas

## 5. SIMULATION AND RESULTS

The performance of 4 QAM modulated data stream is observed under frequency flat AWGN channel condition with channel state information (CSI) known at the receiver. The number of transmitting and receiving antennas selected is (4, 4) .The performance of ZF, MMSE, V-BLAST/ZF and V-BLAST/MMSE are compared with the performance of ML detector.

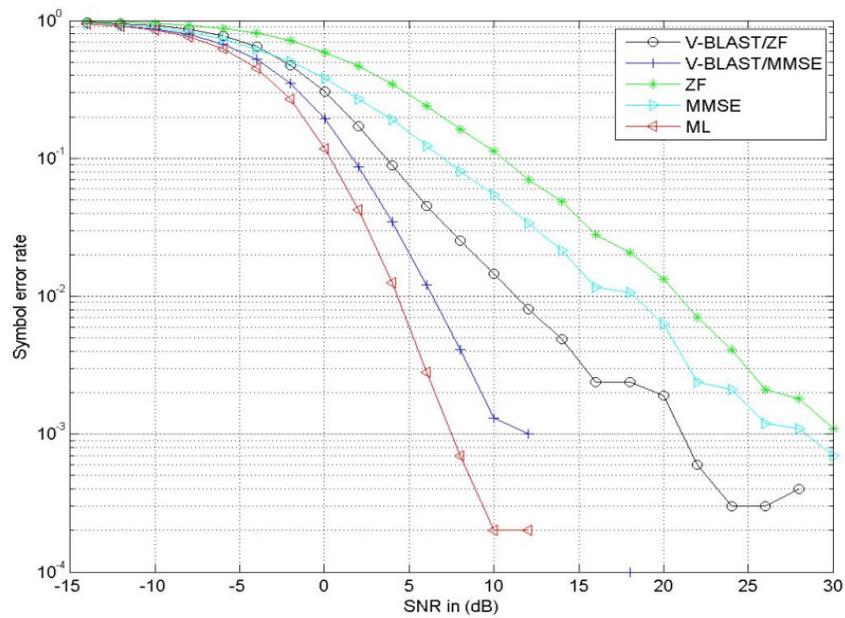

Figure 2. SER plot for 4 QAM modulated symbols under AWGN channel condition with ZF, MMSE, V-BLAST/ZF and V-BLAST/MMSE as symbol detection methods Nt = 4 and Nr = 4

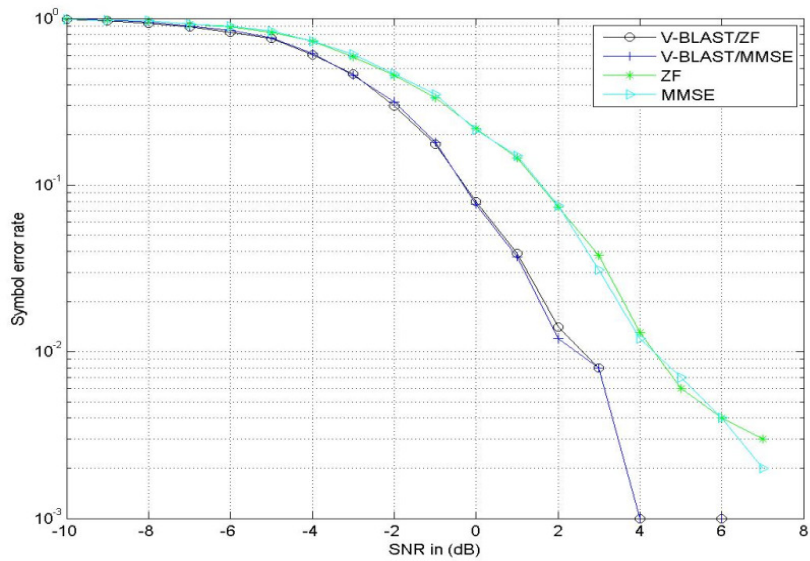

Figure 3. SER plot for 16 QAM modulated symbols under AWGN channel condition with ZF, MMSE, V-BLAST/ZF and V-BLAST/MMSE as symbol detection methods Nt = 6 and Nr = 12

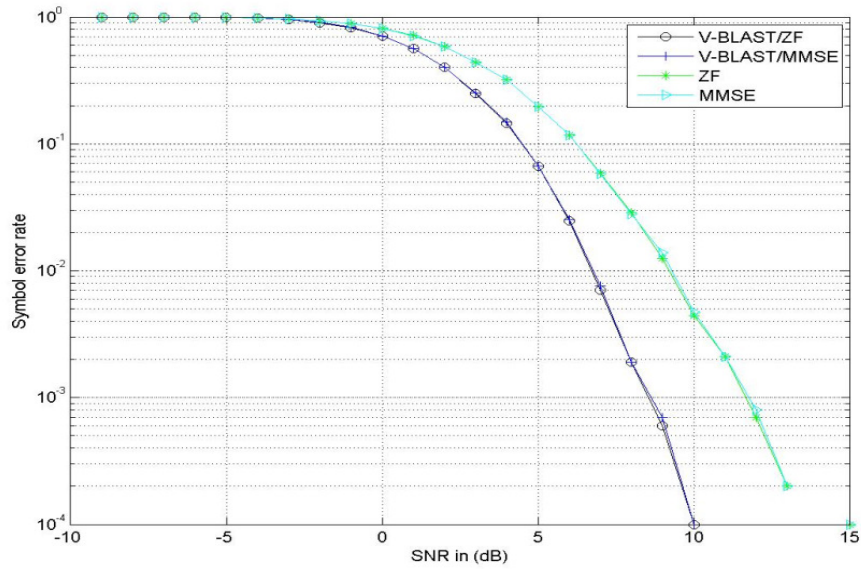

Figure 4. SER plot for 64 QAM modulated symbols under AWGN channel condition with ZF, MMSE, V-BLAST/ZF and V-BLAST/MMSE as symbol detection methods Nt = 6 and Nr = 12

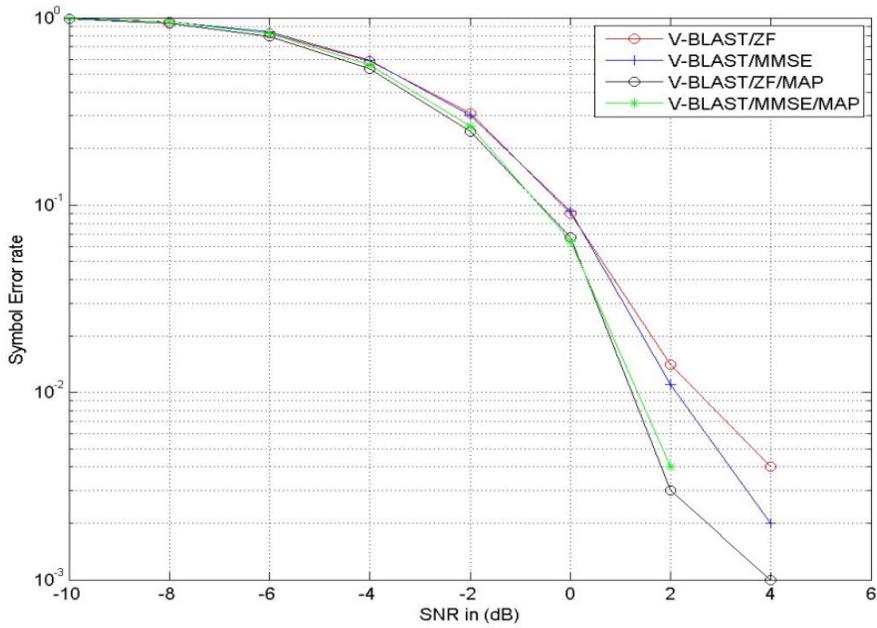

Figure 5. SER plot for 16 QAM modulated symbols under AWGN channel condition with V-BLAST/ZF,V-BLAST/MMSE,V-BLAST/ZF/MAP and VBLAST/ MMSE/MAP as symbol detection methods Nt = 6 and Nr = 12

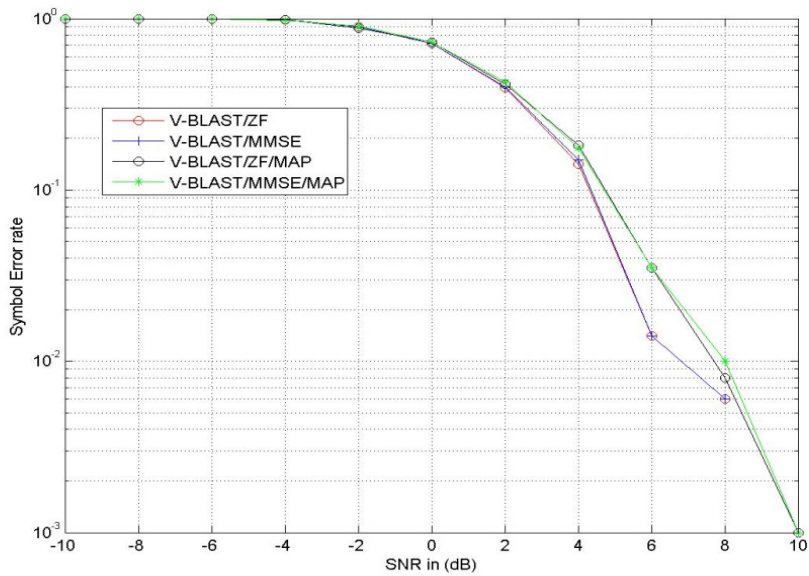

Figure 6.
SER plot for 64 QAM modulated symbols under AWGN channel condition with V-BLAST/ZF,V-BLAST/MMSE,V-BLAST/ZF/MAP and VBLAST/ MMSE/MAP as symbol detection methods Nt = 6 and Nr = 12

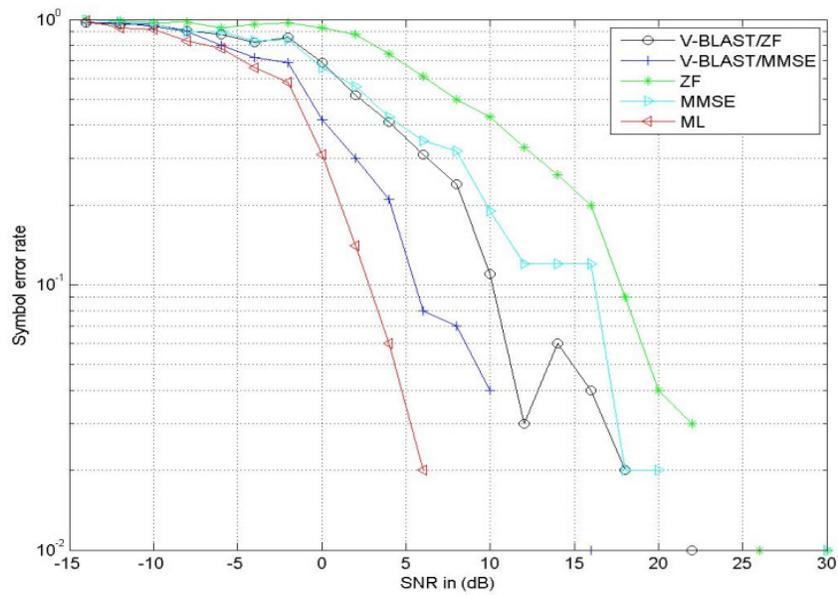

Figure 7. SER plot for 4 QAM modulated symbols under correlated channel condition (ρ = .7) with ZF, MMSE, V-BLAST/ZF and V-BLAST/MMSE as symbol detection methods Nt = 4 and Nr = 4

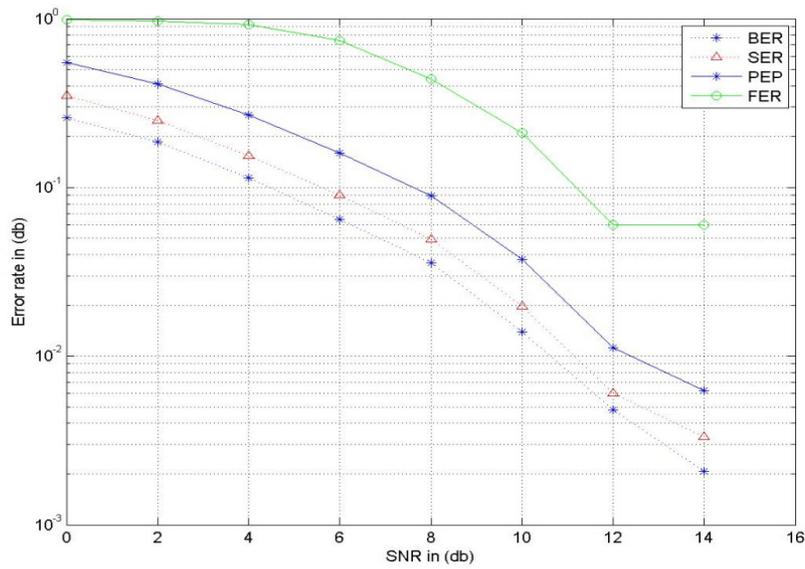

Figure 8. Error rate plot for 4 PSK modulated and STTC data stream under AWGN channel condition with ML as symbol detection methods Nt = 2 and Nr = 2

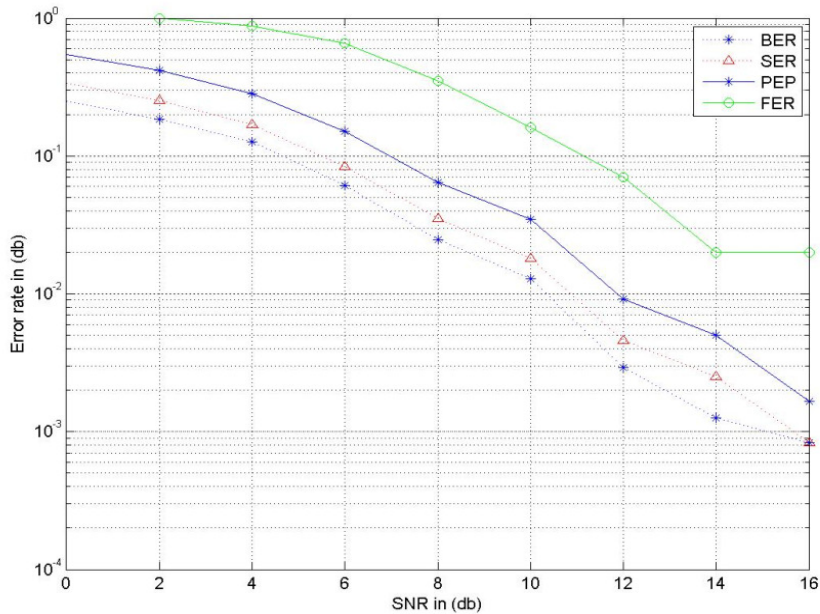

Figure 9. Error rate plot for 4 PSK modulated and STTC data stream under correlated channel condition (ρ = .7) with ML as symbol detection methods Nt = 2 and Nr = 2

The performance of V-BLAST/MMSE is next to it followed by VBLAST/ ZF, MMSE and ZF in the order of decreasing performance. The fig.2 is for 4 QAM constellations. As the constellation size increases the complexity of ML detector increases exponentially. The fig. 3 and fig. 4 show the performance of ZF,MMSE,V-BLAST/ZF and V-BLAST/MMSE for 16 and 64 QAM constellation ,for (6,12) being the number of transmitting and receiving antennas. The performance of V-BLAST/ZF and V-BLAST/MMSE rises when MAP detection algorithm is used. The fig. 5 to fig. 7 compares the performance of V-BLAST/ZF, V-BLAST/MMSE, V-BLAST/ZF/MAP and V-BLAST/MMSE/MAP for the constellation size of 16 and 64. As shown in figure from the fig. 8 and fig.9, symbol error rate plots for V-BLAST/ZF, V-BLAST/MMSE, V-BLAST/ZF/MAP and V-BLAST/ZF/MAP the V-BLAST/ZF/MAP and V-BLAST/MMSE/MAP performance is good for lower constellation size but it degrades for higher constellation size. The performance of ZF, MMSE, V-BLAST/ZF, V-BLAST/MMSE, V-BLAST/ZF/MAP, V-BLAST/MMSE/MAP and ML under correlated channel condition with ρ = .7.The constellation is 4 QAM with Nt = 4 and Nr = 4.

# 6. CONCLUSION AND FUTURE WORK

V-BLAST is a very effective technique for symbol detection which reduces computational complexity manifold but it also works even well in correlated channel condition. To achieve high data rate diversity is exploited in which several replicas of the signal are made available to the receiver in the hope that at least some of them are not attenuated severely. As available bandwidth is finite, the space diversity is promising, since it does not involve any loss of bandwidth. V-BLAST is an example of space diversity scheme. Space time block code and space time lattice code combines both space and time diversity which provides better performance.

In the algorithm implemented assumption was made that all the antenna transmit equal energy but practically depending upon the distance between antennas the transmitted power varies from antenna to antenna. Water filling algorithm is used to estimate the capacity of each path. So, future work can be done by extending these algorithms to system with uneven symbol power under more practical correlated channel condition. Even assumption was made that the CSI is known at the receiver. So, future work involves blind channel estimation of the channel coefficient and assuming that there are errors in channel estimation and then detecting the symbol at receiver when the nature of channel is correlated. The various whitening methods available can use for whitening the channel coefficients.

The various methods for increasing diversity of the data stream are available like STBC, STTC, and LDPC etc. These methods when applied to MIMO OFDM or CDMA system provides high data rate along with good performance. So, V-BLAST technique offers a wide area to work on different systems using various coding scheme under diverse channel condition so that the resultant system has least complexity along with high effectiveness.

## ACKNOWLEDGEMENTS

We would like to express our gratitude to experts Dr. K. K. Chabdda, (Director, PCE), Associate Professor Santosh Choudhary, (HOD, CSE Department), and other members of CSE Dept. for their guidance and contributions. We would also like to thank for the valuable information's they provided us. We would like to thank our family members for their love and care. At last but least we would like to thank everyone, just everyone!

**Authors**

**Shuchi Jani:** completed his B.E. in computer science and engineering from Geetanjali Institute of Technical Studies Udaipur, India. She is pursuing M.tech. from Mewar University and have more than two years teaching experience. She is working as a lecturer in Shrinath Institute of Technology and Engineering (SITE), Nathdwara. Her area of interest for research work includes Wireless Communication, Cryptography and Cloud Computing.

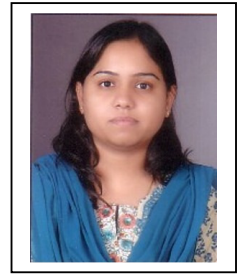

**Shrikrishan Yadav:** working as an Assistant Professor in Computer Science and Engineering Department in PAHER University, Udaipur, India. He has completed B. E. in computer science and engineering from Mohanlal Shukhadia University, Udaipur and pursued M.Tech. in Information Communication from Gyan Vihar University, Jaipur. He has more than two years of experience in academic field. He is also published and presented 9 papers in International and National journals and conferences. He is an associate member of Computer Society of India (CSI), a member of International Association of Engineers (IAENG), International Association of Engineers and Scientist (IAEST) and International Association of Computer Science and Information Technology (IACSIT). His current research interest includes Cognitive Radio, Wireless Sensor Networks, Artificial Intelligence, and Information Communication.

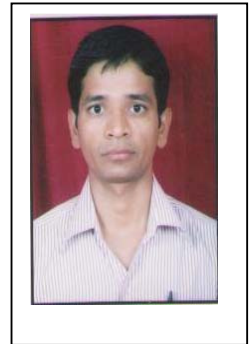

**Bachcha Lal Pal**: working as an Assistant Professor in computer science and information technology department, Mewar University, Chittorgarh, India. He has done B.Tech. in Information Technology from AAI_DU Allahabad U.P. and his M.Tech.(SIT) from DAVV Indore. He is also published some paper in international journals and has attended international conferences. He has more than three years teaching experience. His area of interest for research work includes Wireless Communication, Cryptography, and information technology.

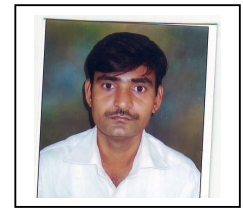